\documentclass[a4paper]{jpconf}
\usepackage{graphicx}
\graphicspath{{fig/}}
\begin{document}
\title{Liquid-gas mixed phase in nuclear matter at finite temperature}

\author{Toshiki Maruyama$^1$ 
and Toshitaka Tatsumi$^2$}

\address{$^1$ Advanced Science Research Center, Japan Atomic Energy Agency, Shirakata Shirane 2-4, Tokai, Ibaraki, 319-1195, Japan}
\address{$^2$ Department of Physics,  Kyoto University, Kyoto, 606-8502 Japan}

\ead{maruyama.toshiki@jaea.go.jp}

\begin{abstract}

We explore the geometrical structure of Liquid-gas (LG) mixed phase 
which is relevant to nuclear matter in the crust region of compact stars or supernovae.
To get the equation of state (EOS) of the system, the Maxwell construction is found to be applicable to symmetric nuclear matter,
where protons and neutrons behave simultaneously.
For asymmetric nuclear matter, on the other hand, the phase equilibrium can be obtained by fully solving 
the Gibbs conditions since the components in the L and G phases are completely different.
We also discuss the effects of surface and the Coulomb interaction on the mixed phase.

\end{abstract}

\section{Introduction}\label{intro}

There are several phase transitions in nuclear matter, some of which
are of the first order.
It is well-known that there appears the mixed phase during the first-order phase transitions (FOPT).
If the system consists of single chemical component, the Maxwell construction
satisfies the Gibbs conditions, i.e.\ balance of pressure and 
chemical potentials among two coexisting phases.
In the case of nuclear matter, however, which consists of several independent
chemical components, one cannot apply the Maxwell construction \cite{gle}.
Due to the Coulomb interaction between charged components in the system, 
the mixed phase shows a series of geometrical structures 
such as droplet, rod, slab, tube and bubble \cite{Rav83}.
They are often called ``pasta structures''  from the rod and slab structures
figuratively spoken as ``spaghetti'' and ``lasagna''.
Appearance of pasta structures is a general feature of the mixed phases 
and common to FOPT in compact stars, such as kaon condensation 
at several times the normal nuclear density \cite{marukaon}, 
hadron-quark (HQ) transition at high density \cite{maru07,maru08} as well as 
liquid-gas phase transition at low-density \cite{marupasta}.

We have developed a framework to study the pasta structure and calculate the equation of state (EOS), 
taking into account the effects of the Coulomb repulsion and the surface tension in a self-consistent way. 
In this scheme the charge screening effects is automatically considered. 
Then we have found that 
the Coulomb screening by the rearrangement of charged particles
reduces the Coulomb energy of the system and consequently
enlarges the size of the structure.
By the screening the local charge density decreases
so that the EOS of the mixed phase approaches 
that of the Maxwell construction.
In particular for the kaon condensation and the HQ transition 
the effects of charge screening are pronounced.
All of the above results are at zero temperature.
For the stellar objects, however,
zero-temperature corresponds only to cold neutron stars.
In contrast, the collapsing stage of supernovae, 
proto neutron stars and neutron star mergers, 
which represent more vivid scene, 
are so warm as several tens MeV.

In this paper we investigate the properties and the EOS of
the mixed phase at finite temperature.
Particularly we are interested in whether there is
any difference between finite and zero-temperature cases.
In the following we concentrate on the low-density 
nuclear matter where the liquid-gas (LG) mixed phase is relevant.

\section{Liquid-gas mixed phase of nuclear matter}

We employ the relativistic mean field (RMF) model to describe the bulk 
properties of nuclear matter under consideration.
The RMF model with fields of 
$\sigma$, $\omega$ and $\rho$ mesons and baryons introduced
in a Lorentz-invariant way is not only relatively simple for 
numerical calculations, but also sufficiently
realistic to reproduce bulk properties of finite nuclei
as well as the saturation properties of nuclear matter \cite{marupasta,marurev}.
One characteristic of our framework is that
the Coulomb interaction is properly included in the 
equations of motion for nucleons and electrons and for meson mean-fields.
Thus the baryon and electron density profiles, as well as the meson
mean-fields, are determined in a fully
consistent way with the Coulomb interaction.

Particle densities are treated within the local-density approximation. 
To solve the equations of motion for the meson mean-fields numerically,
we divided the whole space into equivalent Wigner-Seitz cells with 
geometrical symmetry. 
The shapes of the cell are 
sphere in three-dimensional (3D) case, cylinder in 2D and slab in 1D.
Each cell is globally charge-neutral and all physical quantities
in the cell are smoothly connected to those of the next cell
with zero gradients at the boundary.
The coupled equations for fields in a cell are solved by a relaxation method
for a given baryon-number density 
under constraints of the global charge neutrality.
Parameters included in the RMF model are chosen to reproduce the saturation properties
of symmetric nuclear matter, 
and the binding energies and the proton fractions of nuclei.
Details of the parameters are explained in Refs.\ \cite{marupasta,marurev}.

When we study nuclear matter at finite temperature \cite{yasutake},
the momentum distribution function is a Fermi-Dirac distribution 
instead of a step function.
In the numerical calculation,
density, scalar density, and kinetic energy density, etc of a fermion $a$ 
are obtained by integrating the functions of $T$, $\mu_a$ and $m_a^*$
over all the momentum-space.
We store those values in tables and get necessary quantities by
interpolating them.
The finite-temperature effects for meson excitations and
the contribution of anti-particles are neglected for simplicity.

\subsection{Phase coexistence by a bulk calculation}\label{secBulk}

\begin{figure}
\centerline{
\includegraphics[width=.8\textwidth]{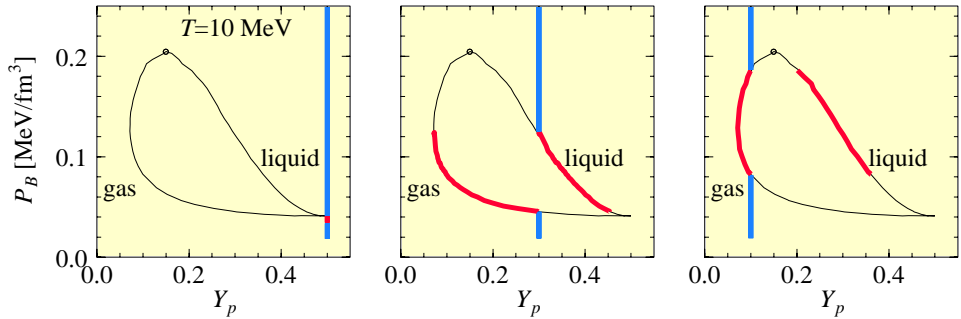}
}
\caption{
Phase coexistence curve on $Y_p$-$P_B$ plane. 
The critical point is denoted by the black circle and the trajectory of the system is 
drawn by the vertical line as pressure is increased.
\vspace{-5mm}
}
\label{figYpP}
\end{figure}

First let us discuss the instability of uniform matter by the use of the bulk calculation, 
where free electrons are included  to neutralize the whole system \cite{cho}.
Here, pressure balance and chemical equilibrium of fermions 
are imposed between coexisting two uniform phases.
The Coulomb interaction and gradient terms of meson mean-fields 
are then ignored.
Figure \ref{figYpP} shows phase coexistence (binodal) curves 
on the $Y_p$-$P_B$ plane for given temperatures, 
where $Y_p$ is the proton ratio and
$P_B$ is the baryon partial pressure.
The left part of the thin solid curves corresponds to
the neutron-rich phase of the coexisting phases and vice versa.
Since the symmetry energy is positive in our model,
nuclear matter with smaller proton ratio has a smaller 
density at a given pressure. 
Each curve is divided into two regions by the critical point $y_{\rm cp}$, 
which is defined by $dP_B/dy|_{y_{\rm cp}}=0$, and each region corresponds to the  gas or liquid phase.
Roughly speaking, the high-pressure region  
is in the liquid phase, and the low-pressure region the gas phase. 
Two phases cannot be distinguished at the critical point.
The interior of the curves is a forbidden region.
Consider a compression of nuclear matter with fixed $Y_p=0.3$ in 
the middle panel of Fig.\ \ref{figYpP}.
Starting from the bottom, the baryon pressure goes up
until $P_B\approx 0.04\ {\rm MeV/fm^3}$.
Then it encounters the coexistence curve.
The system cannot enter the interior of the coexistence curve
and causes the phase separation into gas (dilute) and liquid (dense) phases 
with different values of $Y_p$; the phase with larger value of $Y_p$ is the liquid phase, 
while the phase with smaller value of $Y_p$ is the gas phase. 
This is due to the positive symmetry energy.
With increase of the pressure or the density,
the volume fraction of the liquid phase increases from $0$ to $1$ until 
$Y_p$ of the liquid phase becomes 0.3.
At this point, all the system is occupied by the liquid phase
and the gas phase vanishes.

In the case of $Y_p=0.5$ in the left panel of Fig.\ \ref{figYpP},
The trajectory of the system and the coexistence curve 
meet at a single point on the $Y_p$-$P_B$ plane.
Therefore the mixed phase of this system consists of
gas and liquid with the same value of $Y_p=0.5$.
This means that the baryon system behaves as that with
a single component.
In other words the phase transition is ``congruent'' \cite{refIosilevskiy}.
As briefly discussed in Sec.\ \ref{intro},
the EOS of such a system can be obtained by the Maxwell construction.
In fact one can also see that during the phase transition from
gas to liquid, the baryon partial pressure $P_B$ remains constant 
at $\approx 0.04\ {\rm MeV/fm^3}$.
However, one should keep in mind that the above argument applies to
the baryon partial system, not the total system including electrons.
The total pressure is dominated by electrons and 
a monotonic function of density. 
One cannot apply the Maxwell construction to the total pressure.

\begin{figure}[b]
\centerline{
\includegraphics[width=.45\textwidth]{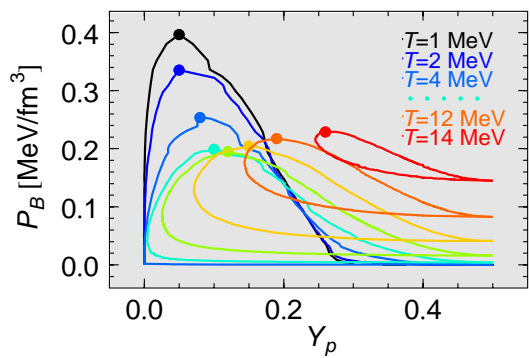}
}
\caption{
Temperature-dependence of phase coexistence curves for given temperatures. 
Filled circles show the critical points.
}
\label{figYP}
\end{figure}

\begin{figure}
\centerline{
\includegraphics[width=.8\textwidth]{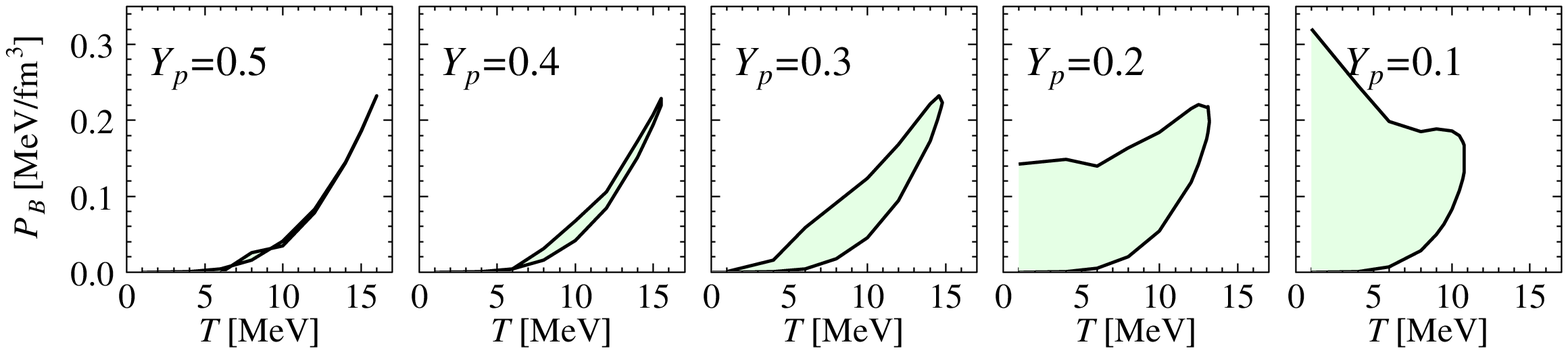}
}
\caption{
Phase coexistence curve on $T$-$P_B$ plane. 
Pale green shows the region of LG mixed phase.
}
\label{figTP}
\end{figure}

In the case of $Y_p=0.1$, the system starts in a gas phase  
at the bottom of the right panel of Fig.\ \ref{figYpP}.
With increase of density, the pressure goes up till
$P_B\approx0.08\ {\rm MeV/fm^3}$.
Then the LG mixed phase appears as in the middle panel. 
With further increase of density, the pressure goes up
until $P_B\approx0.18\ {\rm MeV/fm^3}$.
At this point $Y_p$ of the gas phase becomes that of 
the total system, 0.1, which means that the system
becomes a gas again. However such ``supercritical'' 
gas has very different properties from the usual gas at low pressure.
Such a transition, (gas) $\rightarrow$ (mixed phase) $\rightarrow$ (supercritical gas),
is peculiar to systems with two or more chemical components
and called ``retrograde condensation'' \cite{refRetro}.

In Fig.\ \ref{figYP} we show the temperature-dependence of the 
phase coexistence curves.
One can see that at higher temperature the
region of mixed phase are limited and the
congruence is enhanced, 
i.e., the difference in $Y_p$ of coexisting phases becomes small.

Figure \ref{figTP} shows the phase coexistence region on $T$-$P_B$ plane. 
The region with pale green indicate mixed phase.
One can see that at $Y_p=0.5$, the phase transition occurs 
below the critical temperature $\sim 16$ MeV and it is congruent; EOS exhibits a constant-pressure region as the Maxwell construction leads.
With increase of $Y_p$, the region of the mixed phase in $T$-$P_B$ spreads,
which shows the increase of non-congruence \cite{refIosilevskiy}. 

\subsection{Phase transition with ``pasta'' structures}

\begin{figure}[b]
\centerline{
\includegraphics[width=.28\textwidth]{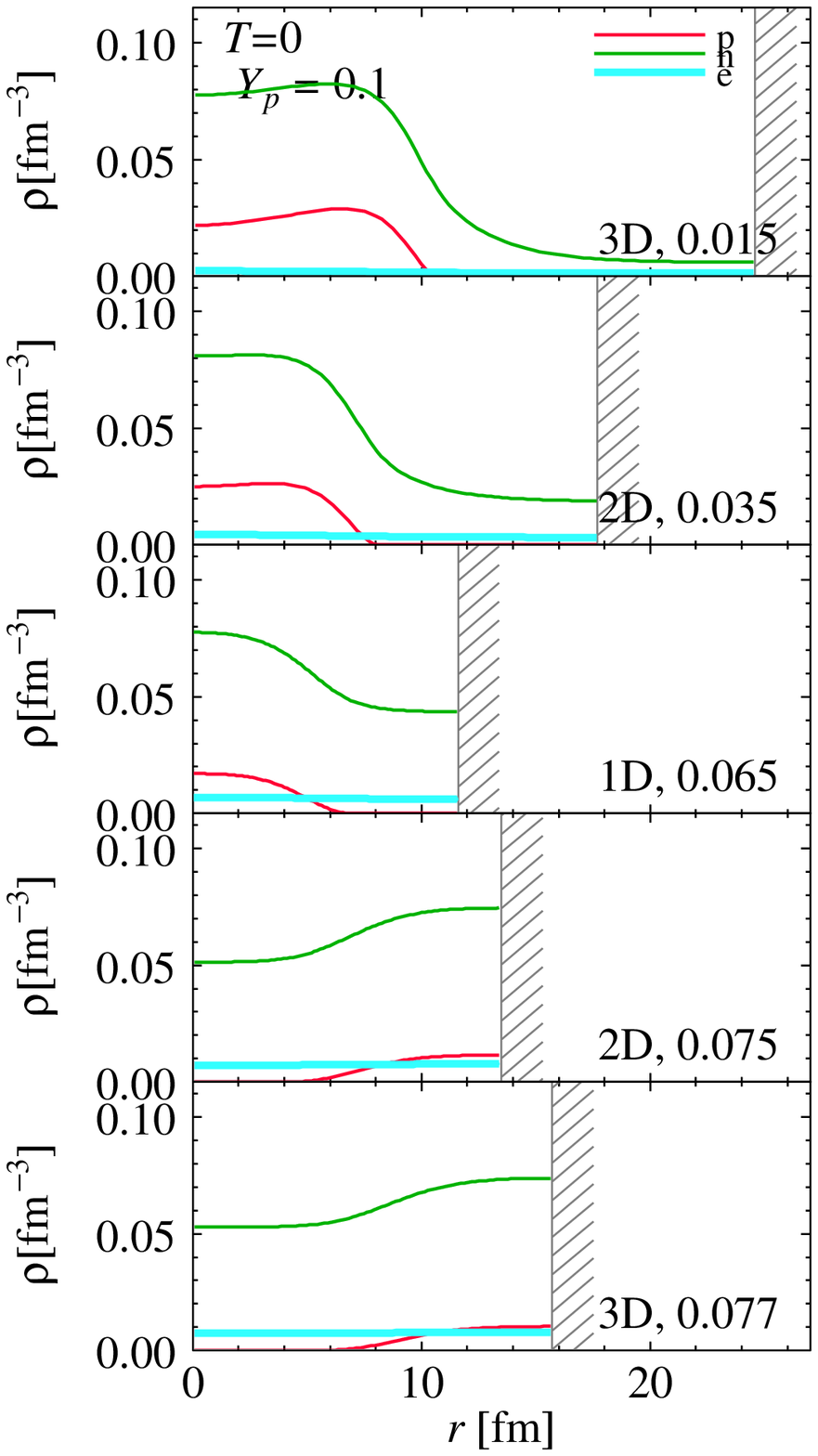}
\includegraphics[width=.28\textwidth]{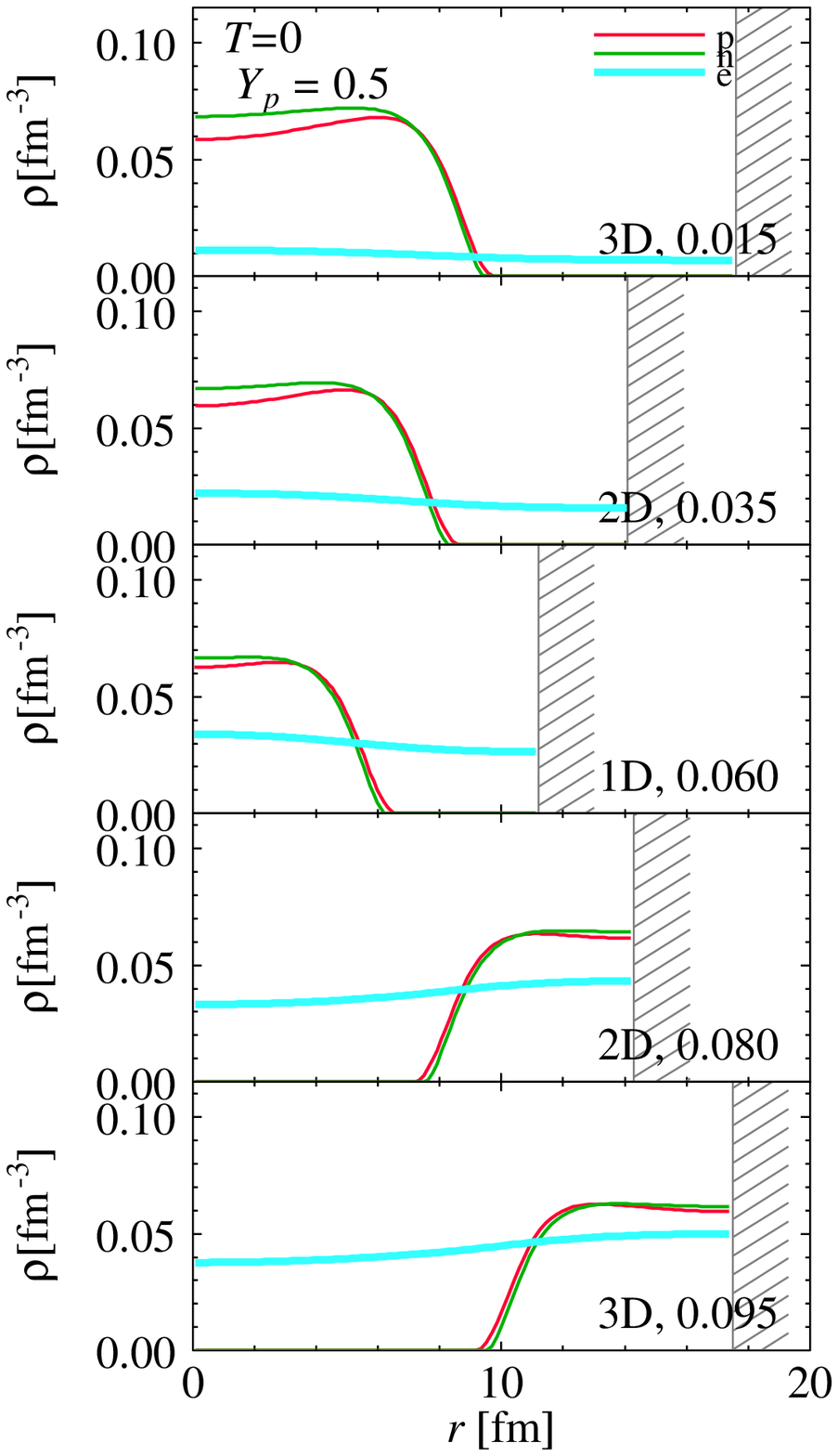}
\includegraphics[width=.28\textwidth]{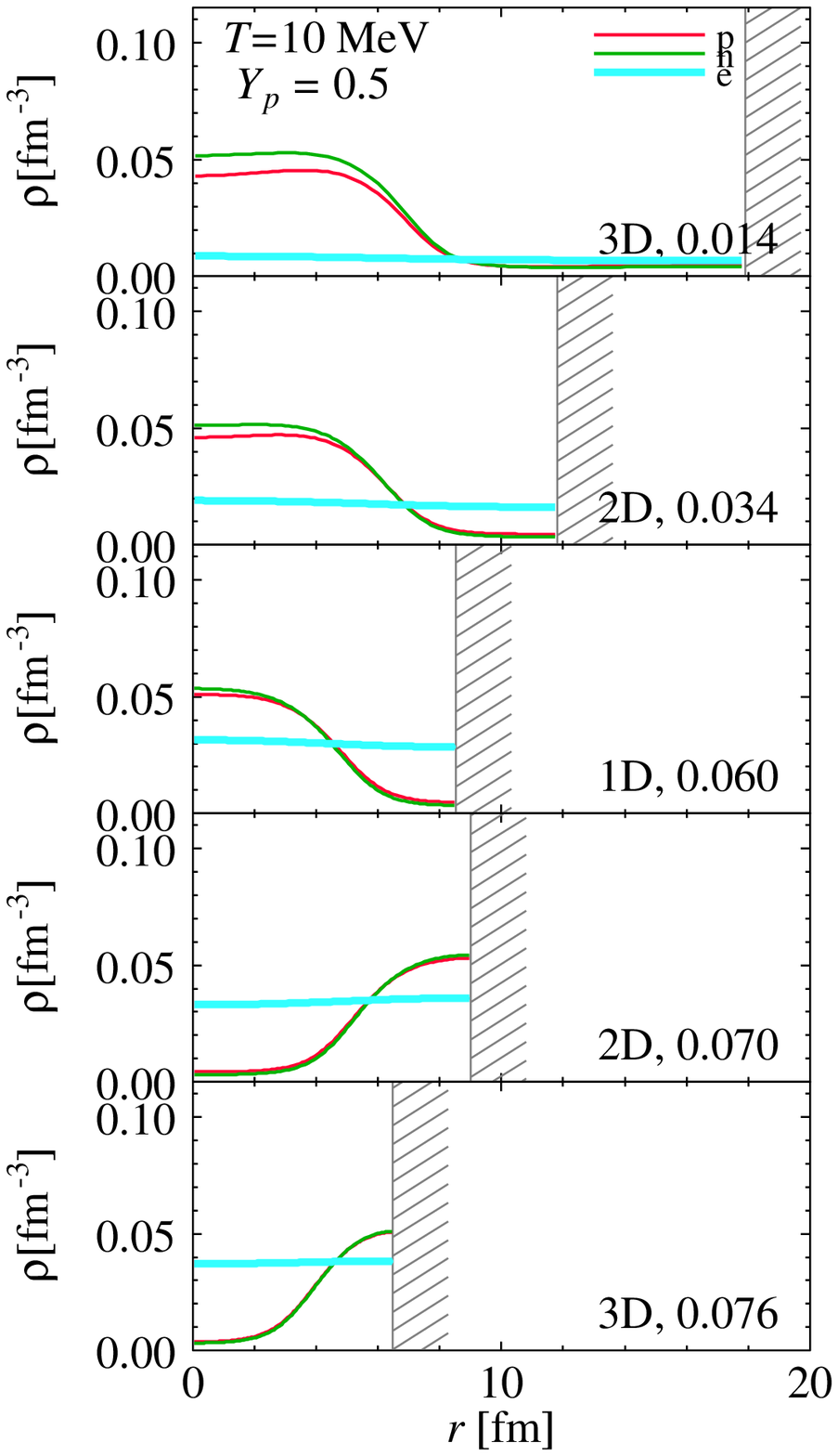}
}
\caption{
Examples of the density profiles in the Wigner-Seitz cells.
with $Y_p=0.1$ (left), $Y_p=0.5$ $T=0$ (middle) and  $T=10$ MeV (right).
}
\label{figPasta}
\end{figure}

Next, we discuss the properties of nuclear matter with pasta structures.
In this case we fully solve equations of motion for meson mean fields
with derivative terms, which supply surface energy of non-uniform
baryon distribution.
The Coulomb interaction among charged particles is taken into account.
The cell size and the geometrical shape are searched for 
so that the free energy density becomes minimum.

We show in Fig.~\ref{figPasta} some typical density
profiles in the cell. 
The left and the middle panels show the cases of proton fraction $Y_p=0.1$ and $Y_p=0.5$.
Apparently, dense nuclear phase (liquid) and dilute nuclear/electron phase (gas) 
are separated in space and they form pasta structures depending on density.
One should notice that coexisting two phases 
are non-congruent and have different components, 
i.e.\ nuclear matter and electron gas.
Therefore the EOS of the whole system cannot be obtained
by the Maxwell construction.
Since electron density is almost uniform 
and independent of baryon density distribution,
we can separately discuss the properties of the baryon partial system.

\begin{figure}
\centerline{
\includegraphics[width=.42\textwidth]{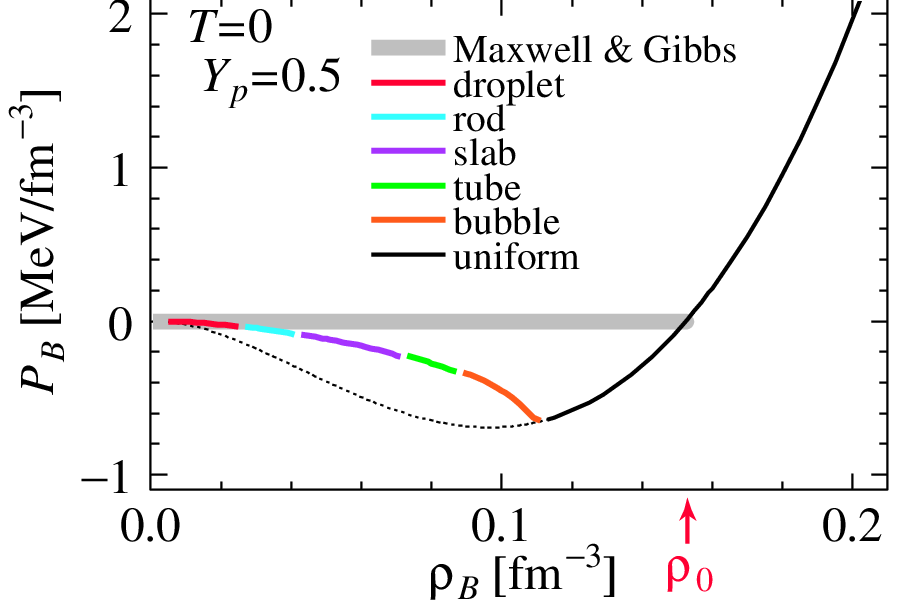}
\includegraphics[width=.42\textwidth]{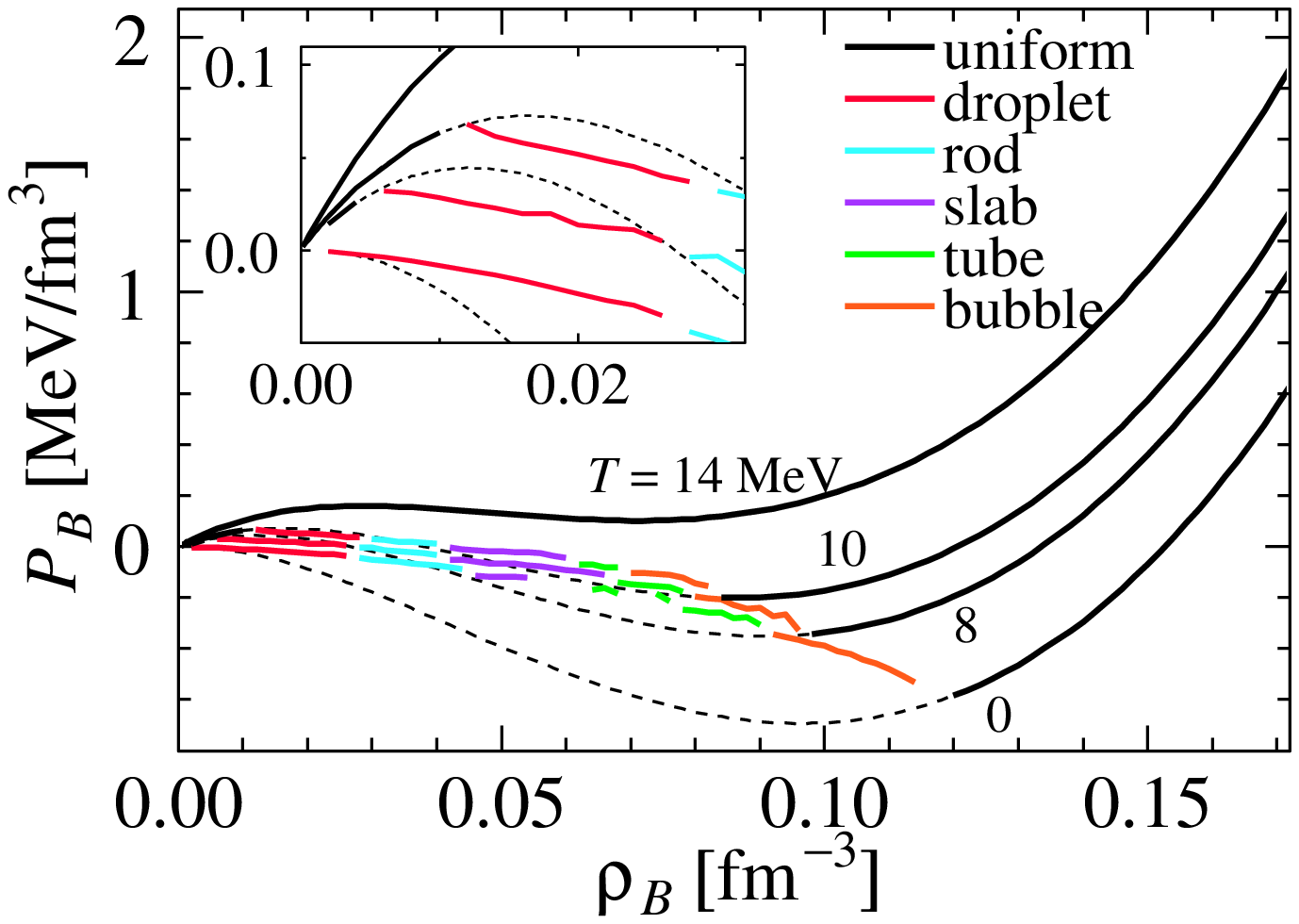}
}
\caption{
Baryon partial pressure as a function of density $\rho_B$
for symmetric nuclear matter $Y_p=0.5$, 
which is given by subtracting electrons pressure from the total one.
\vspace{-5mm}}
\label{figEOS05}
\end{figure}

In the case of $Y_p=0.5$, as discussed in Sec.\ \ref{secBulk},
the local proton fraction stays almost constant.
Therefore the system behaves like a system with single component.
This means that one can apply the Maxwell construction to get 
the baryon partial pressure $P_B$ as in the left panel of Fig.~\ref{figEOS05}:
uniform low-density matter with a negative partial pressure is 
not favored and the Maxwell construction gives $P_B=0$ for the mixed phase. 
By the finite-size effects, i.e.\ the Coulomb repulsion and the surface tension,
the structured mixed phase becomes unstable in the density region 
just below $\rho_0$,
and consequently uniform matter with a negative partial pressure is allowed. 
Note again that the total pressure including electrons is always positive 
even in this case, so that the system is thermodynamically stable.

In the case of asymmetric nuclear matter, 
e.g.\ $Y_p=0.1$ in the left panel of Fig.~\ref{figPasta},
the proton fraction in the dilute and dense phases are 
different, especially for low $Y_p$.
Matter behaves as a system with multi chemical components
and the Maxwell construction does not satisfy the Gibbs conditions.

\subsection{Thermal effects}

Let us discuss the thermal effects on the 
LG mixed phase of low density nuclear matter.
By comparing the density profiles in the middle and right panels
of Fig.~\ref{figPasta},
we easily notice that 
the dilute phase at finite temperature always contains baryons 
while they are absent in the dilute phase at zero temperature if $Y_p\approx 0.5$.
This is due to the Fermi distribution at finite temperature, where 
density as a function of chemical potential is always positive.

We also notice that the size of the pasta structure is smaller
in the case of finite temperature.
This comes from a reduction of the surface tension between two phases
at finite temperature since the difference of baryon densities  
between two phases get smaller.

The isothermal EOS's (baryon partial pressure as a function of baryon number density)
of symmetric nuclear matter 
at various temperatures are shown in the left panel of Fig.~\ref{figEOS05}.
Dotted and thick solid curves show the cases of uniform matter,
while thin solid curves are the cases where non-uniform pasta structures 
are present.
As shown in the right panel of Fig.~\ref{figEOS05},
pasta structures appear at finite temperatures as well as the case of $T=0$.
But there appears uniform matter (gas phase) at the lowest-density region
\cite{avancini,friedman}
since the baryon partial pressure of uniform matter has
a positive gradient against density.
On the other hand, the uniform matter is unstable
where the pressure gradient is negative
even if the pressure itself is positive.
At $T=14$ MeV, we obtain no pasta structure
since the baryon partial pressure of uniform matter
becomes a monotonic function of density above this temperature.
Study of the instability of uniform matter and the appearance of the pasta structures  
in connection with the spinodal region is in progress. 

In Sec.\ \ref{secBulk} we have shown a retrograde condensation
at $T=10$ MeV and $Y_p=0.1$ by the bulk calculation.
In the full calculation with pasta structures, however, 
there is no evidence to have such phenomenon so far.
Probably it is washed away by the finite-size effects.

\section{Summary and concluding remarks}

We have investigated the properties of LG mixed phase in low-density nuclear matter.
We have seen that pasta structures appear at finite temperatures as well as
at zero temperature.
For the LG mixed phase, the proton fraction $Y_p$ is a crucial quantity.
If $Y_p\approx0.5$ the local proton fraction is almost constant and
baryon partial system behaves like a system with a single component.
Therefore the Maxwell construction is applicable in the bulk calculation. 
However, when pasta structures are considered
the finite-size effects reduce the region of the mixed phase
by the mechanical instability of pasta structures.
At finite temperatures the size of the pasta structures becomes smaller.
This is due to the reduction of the surface tension 
between liquid and gas phases.

In conclusion we emphasize that the existence of
pasta structures together with the 
finite-size effects are
common and general for the mixed phases at 
the FOPT in nuclear matter.
Particularly matter at finite temperature
exhibits various features which are
interesting and important for 
not only nuclear astrophysics but also thermodynamics.

\ack
We would like to thank Prof.\ I.~Iosilevskiy,
Dr.\ N.~Yasutake and Dr.\ S.~Chiba
for fruitful discussions.

\end{document}